\documentclass{aa}

\usepackage{amsmath, amsthm, amssymb, amsfonts}
\usepackage{graphicx}
\usepackage{natbib}
\usepackage{subfig}
\usepackage{siunitx}
\usepackage{cleveref}
\usepackage{xcolor}
\usepackage[T1]{fontenc}
\crefname{section}{§}{§§}
\Crefname{section}{§}{§§}
\crefformat{section}{§#2#1#3}
\bibpunct{(}{)}{;}{a}{}{,} 

\begin{document}
        
        \title{Analysing the structure of the bulge with Mira variables}
        
        \subtitle{}
        
        \author{\v{Z}. Chrob\'{a}kov\'{a}\inst{1,2}, M. L\'{o}pez-Corredoira\inst{1,2}, F. Garz\'{o}n\inst{1,2}}
        
        \institute{Instituto de Astrofísica de Canarias, E-38205 La Laguna, Tenerife, Spain
                \and
                Departamento de Astrofísica, Universidad de La Laguna, E-38206 La Laguna, Tenerife, Spain}
        
        \date{Received xxxx; accepted xxxx}
        
        
        \abstract
        {The Galactic bulge at latitude $4<|b|$(deg.)$<10$ was claimed to show an X-shape, which means that stellar density distributions along the line of sight have a double peak. However, this double peak is only observed with the red-clump population, and doubt has been cast on its use as a perfect standard candle.\ As such, a boxy bulge without an X-shape is not discarded.}
        {We aim to constrain the shape of the bulge making use of a different population: Mira variables from the new Optical Gravitational Lensing Experiment data release, OGLE-IV, with an average age of 9 Gyr.}
        {We analysed an area of the bulge far from the plane, where we fitted the density of the Miras with boxy bulge and X-shaped bulge models and calculated the probability of each model.}
        {We find that the probability of a boxy bulge fitting the data is $p=0.19$, whereas the  probability for the X-shaped bulge is only $p=2.85 \cdot 10^{-6}$ (equivalent to a tension of the model with the data of a 4.7$\sigma $ level). Therefore, the boxy bulge model seems to be more appropriate for describing the Galactic bulge, although we cannot exclude any model with complete certainty.}
        {}
        
        \titlerunning{Analyzing the structure of the bulge with Mira variables}
        \authorrunning{\v{Z}. Chrob\'{a}kov\'{a} et al.}
        \maketitle
        
\section{Introduction}  
The Galactic bulge morphology is an intensely discussed topic. The first near-infrared surveys discovered it to be non-axisymmetric \citep{weiland,martin97}, but its shape has been debated. \cite{weiland} showed the bulge to have a peanut shape  using the Cosmic Background Explorer survey. This peanut shape was later interpreted as the imprint of a boxy bulge \citep{kent,dwek,martin05} due to a composite effect expected to appear considering the stable orbits of several families of periodic orbits \citep[e.g.,][]{pastis}.

More recently, the bulge has been reported to have an X-shape, based on an analysis involving red-clump stars \citep[e.g.,][]{mcwilliam,wegg,nataf}. These studies find a double peak in the star counts along the line of sight, which leads to an X-shape in the density. However, \cite{martin16} and \cite{Lop19} argue that the apparent X-shape could be an artefact and that the red-clump stars do not have a unique narrow peak in their luminosity function. Further doubt was cast by \cite{lee}, who attribute the double peak in bulge density to an effect of multiple populations being present in the bulge, although \cite{gonzalez} refute this explanation. The X-shaped bulge was also observed using infrared images of the Milky Way, for example by \cite{ness}. However, these images are dependent on the image processing with the subtraction of some particular disk model or there may be artefacts from subtracting the bulge as an ellipsoid instead of as a boxy bulge \citep{han}.

\begin{figure*}
	\centering
	\includegraphics[width=0.35\textwidth]{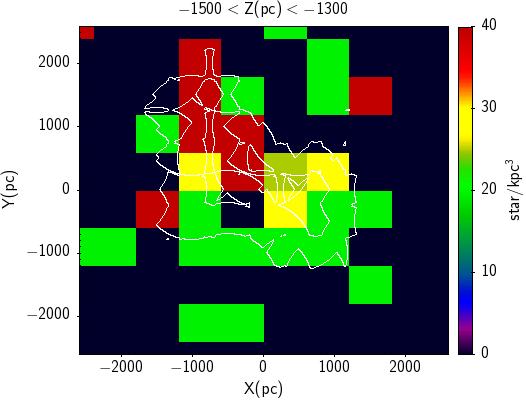}
	\includegraphics[width=0.35\textwidth]{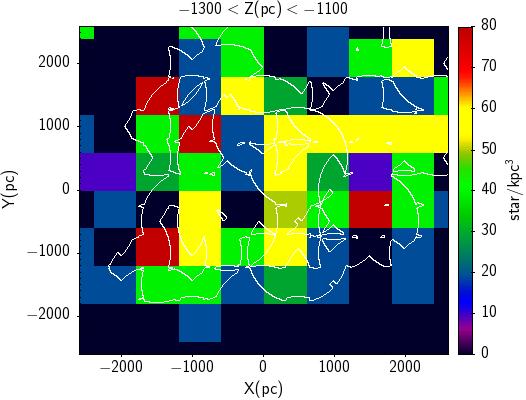}
	\includegraphics[width=0.35\textwidth]{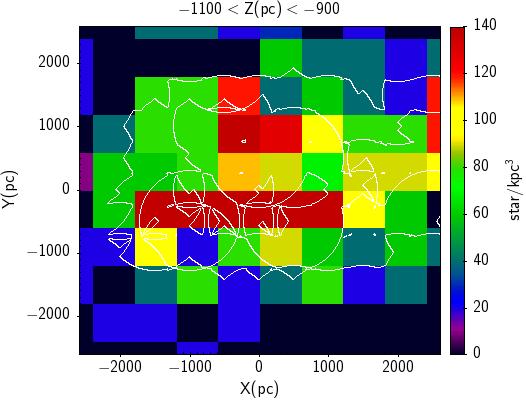}
	\includegraphics[width=0.35\textwidth]{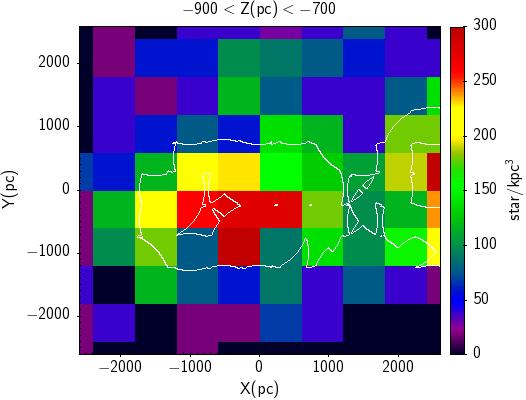}
	\includegraphics[width=0.35\textwidth]{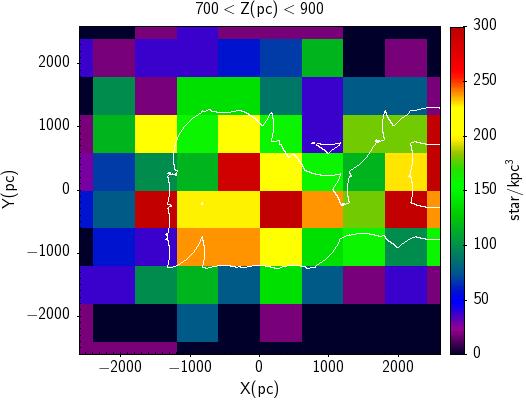}	\includegraphics[width=0.35\textwidth]{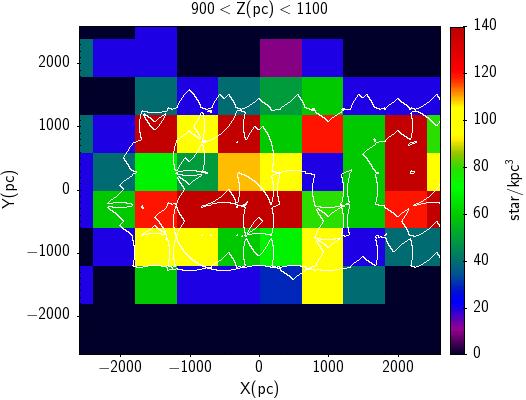}
	\includegraphics[width=0.35\textwidth]{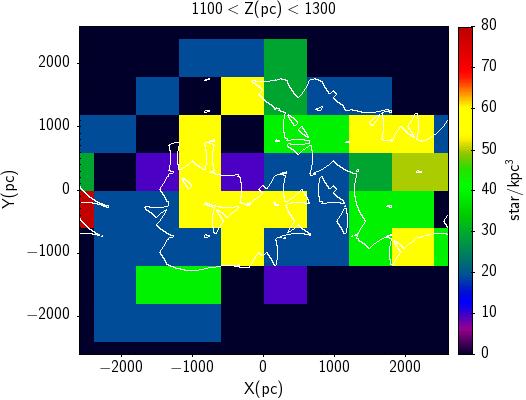}
	\includegraphics[width=0.35\textwidth]{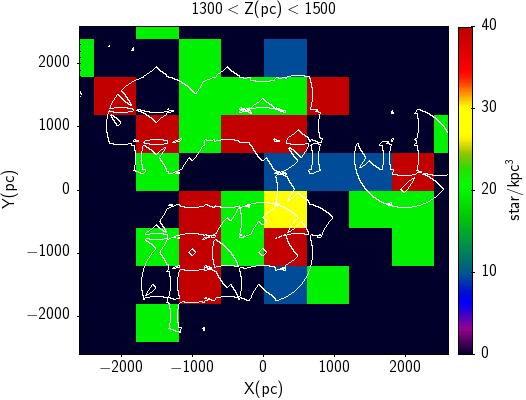}
	\caption{Density, $\rho $, of Mira variable stars as a function of galactocentric coordinates (the X axis is in the Sun-Galactic centre direction such that $X_\odot =+8.2$ kpc) in the ranges  $|X|\le 2750$ pc, $|Y|\le 2750$ pc, and 700 pc$<|Z|<$1500 pc. Bin sizes are 500x500x200 pc, which implies a Poissonian error of the density in each bin equal to $\Delta \rho =\sqrt{\rho \rho_1}$ with $\rho _1=20\ {\rm star/kpc}^3$; for $\rho =0$, $\Delta \rho =\rho _1$. Over-imposed white contours show isodensity regions.}\label{Fig:slices}
\end{figure*}

In populations other than red-clump populations, the X-shaped bulge is not observed. For instance, very old RR Lyrae stars \citep{dekany,pietrukowicz}, young populations ($\lesssim 5$ Gyr)   such as F0-F5V stars \citep{martin16}, and Mira variables of all ages \citep{Lop17} 
do not manifest an X-shaped bulge. \cite{vasquez}, \cite{ness_2014}, and \cite{rojas} presented evidence that the X-shape bulge of the Milky Way is only traced by the metal-rich bulge stars. Recently, \cite{semczuk} analysed the morphology of the bulge with Miras of various ages and found an age-morphology dependence consistent with a boxy/peanut bulge.

Mira variables are pulsating stars, with periods ranging from about 80 to over 1000 days. They are cool giant stars near the tip of the asymptotic giant branch with a high brightness and a well-defined period-luminosity relation, which makes them excellent distance indicators and a useful population for studying Galactic structure \citep{Iwa22}. In this Letter we repeat the analysis by \cite{Lop17} for the density of the bulge with Mira variables, but with a much larger coverage and number of stars provided by the recent data release of the Optical Gravitational Lensing Experiment survey (OGLE-IV). 

The Letter is structured as follows. In Sect. 2 we describe the data selection. In Sect. 3 we derive the density distribution and fit it with models of boxy and X-shaped bulges. In Sect. 4 we conclude.

\section{Data}
We used OGLE-IV data of Mira variables \citep{Iwa22}, carried out with the 1.3 m Warsaw Telescope at the Las Campanas Observatory in Chile. The catalogue covers the whole Galactic bulge area with 40\,356 objects, most of them near the plane, and contains an additional 25\,625 stars in the Galactic disk, observed in the Johnson $V$-band (mean wavelength of \SI{0.55}{\micro\metre}) and Cousins $I$-band (mean wavelength of \SI{0.81}{\micro\metre}) filters. 

We analysed the bulge using stars far from the plane (700 pc$\le |Z|<1500$ pc) in order to avoid the problems of incompleteness due to extinction, and also because the X-shape features are only notable within this range of $Z$. 
The completeness is estimated to be 96\% \citep{Iwa22}; $\sim 4$\% of Miras are not classified as such due to the small number of epochs.

\begin{figure*}
	\centering
	\includegraphics[width=0.35\textwidth]{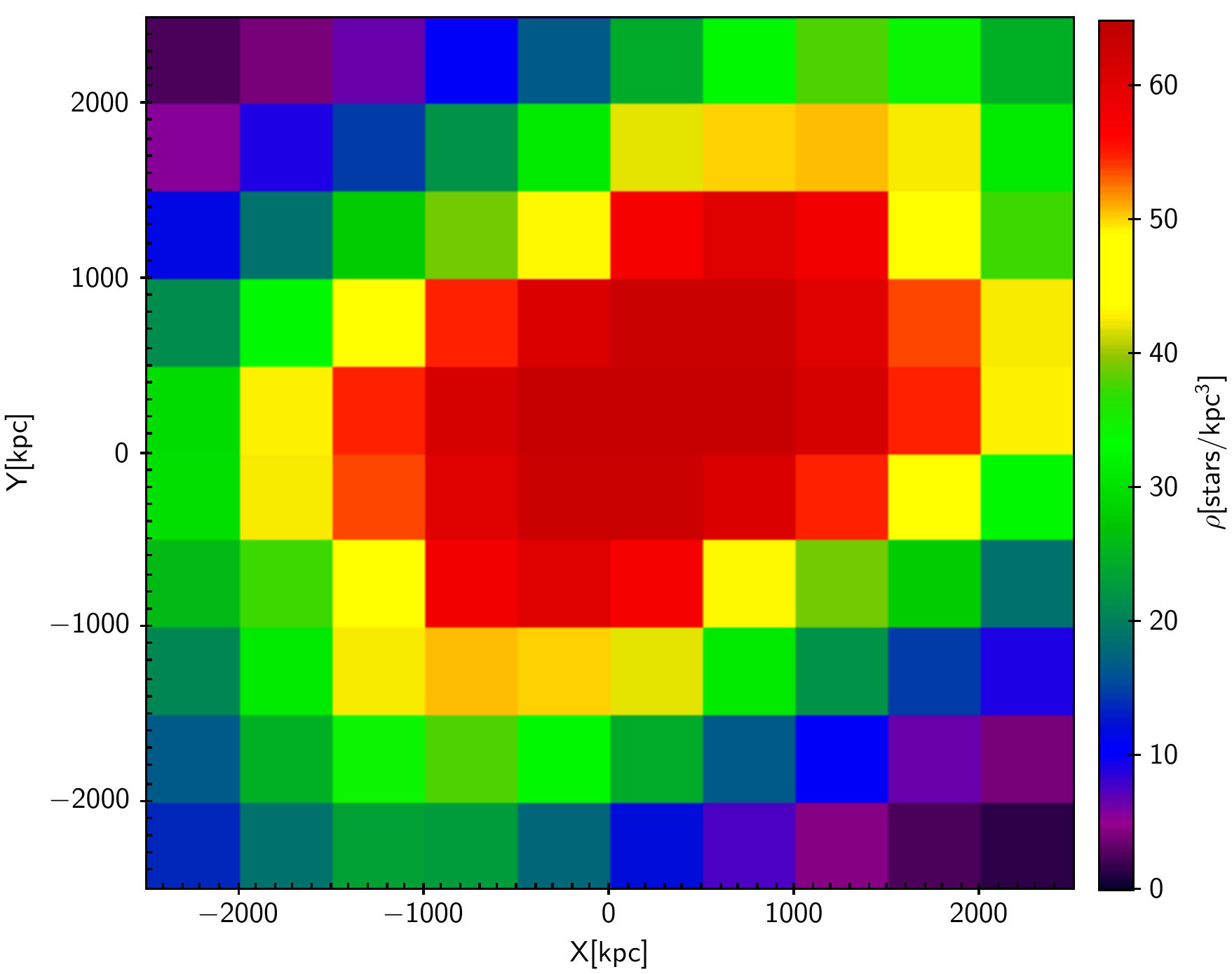}
	\includegraphics[width=0.35\textwidth]{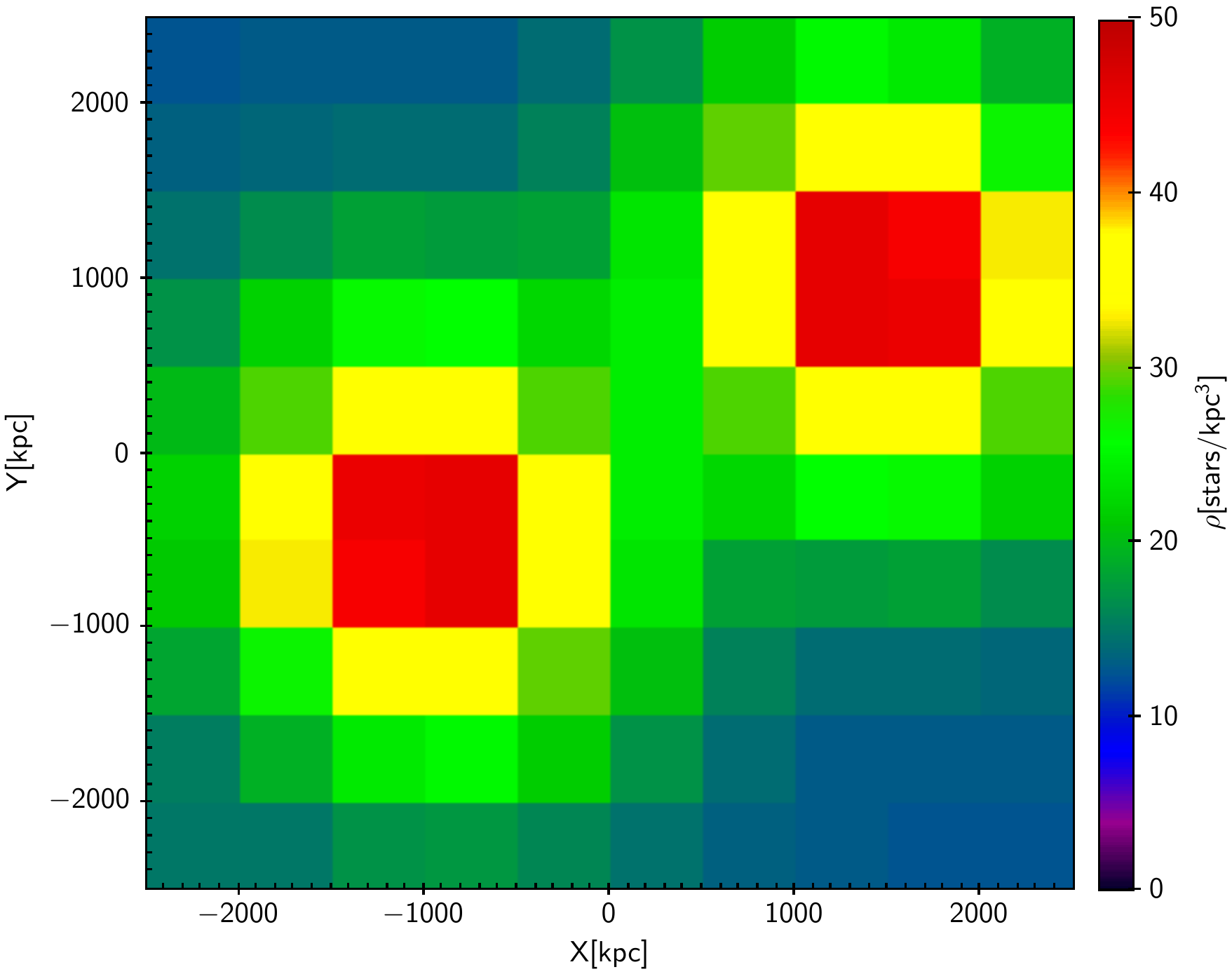}
	\includegraphics[width=0.35\textwidth]{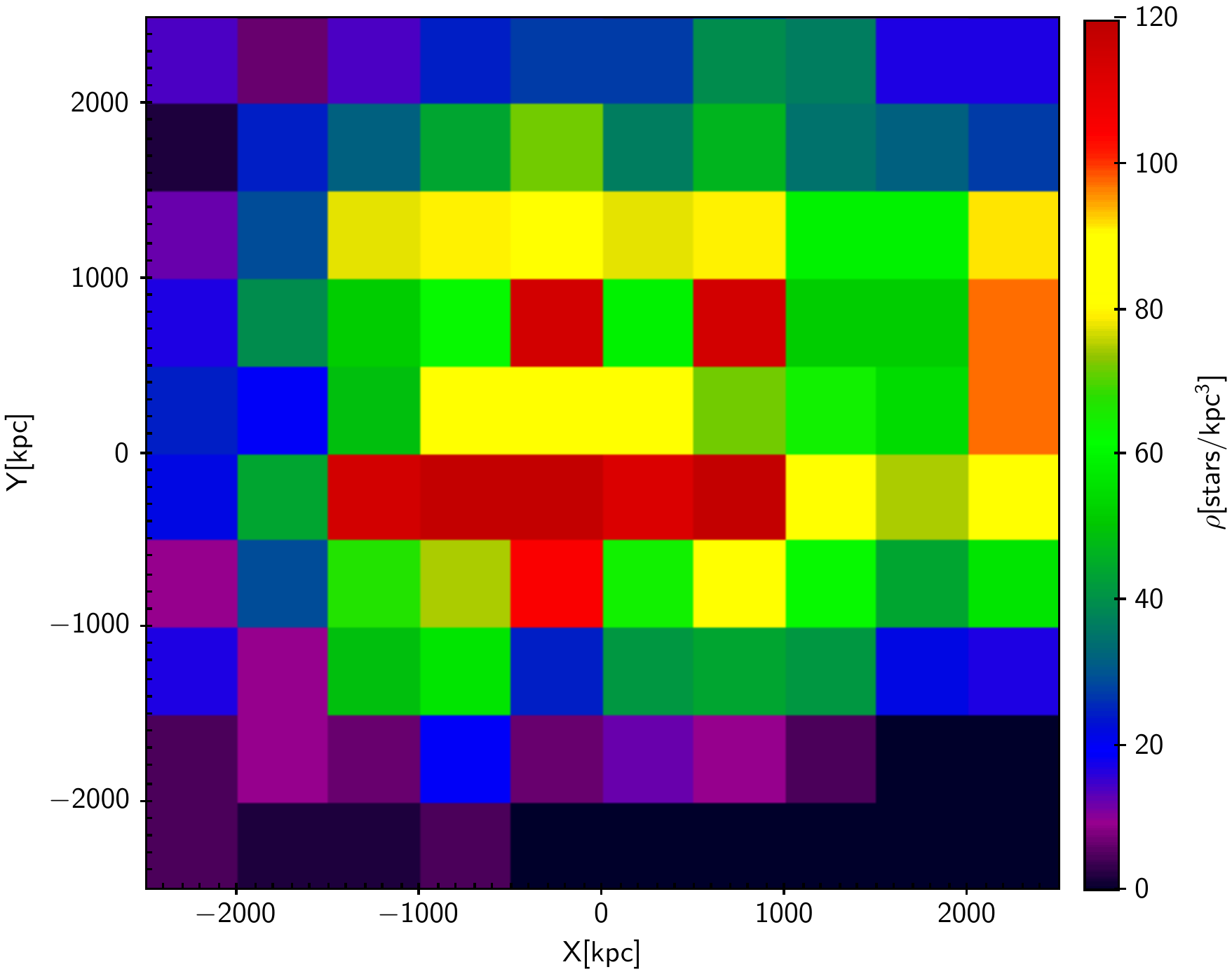}
	\caption{Models of the density averaged for all bins of $Z$ that we use compared with the data. (a): Density of the box bulge. (b): Density of the X-shaped bulge. (c): Data.}\label{dens}
\end{figure*}

\begin{figure*}
	\centering
	\includegraphics[width=0.35\textwidth]{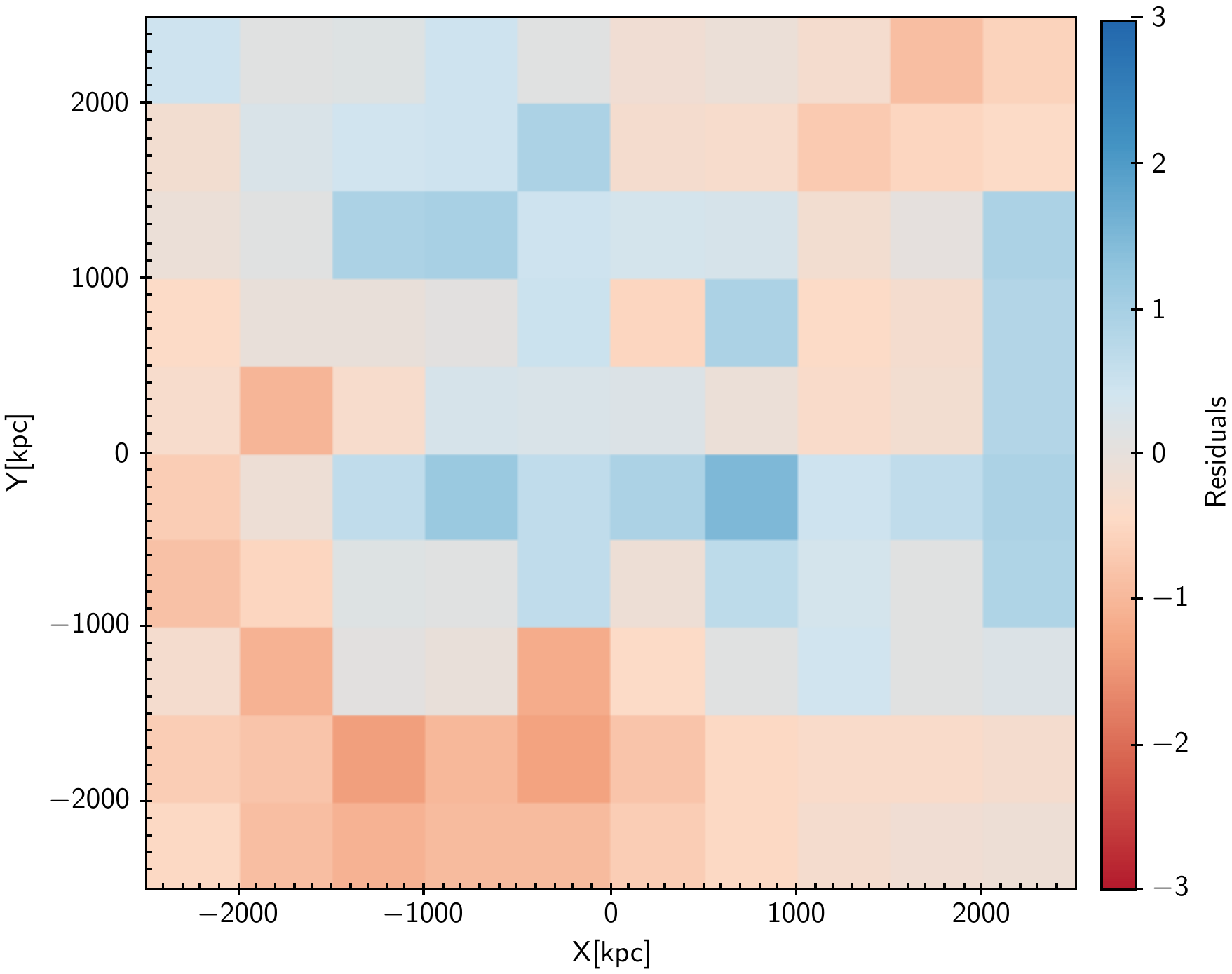}
	\includegraphics[width=0.35\textwidth]{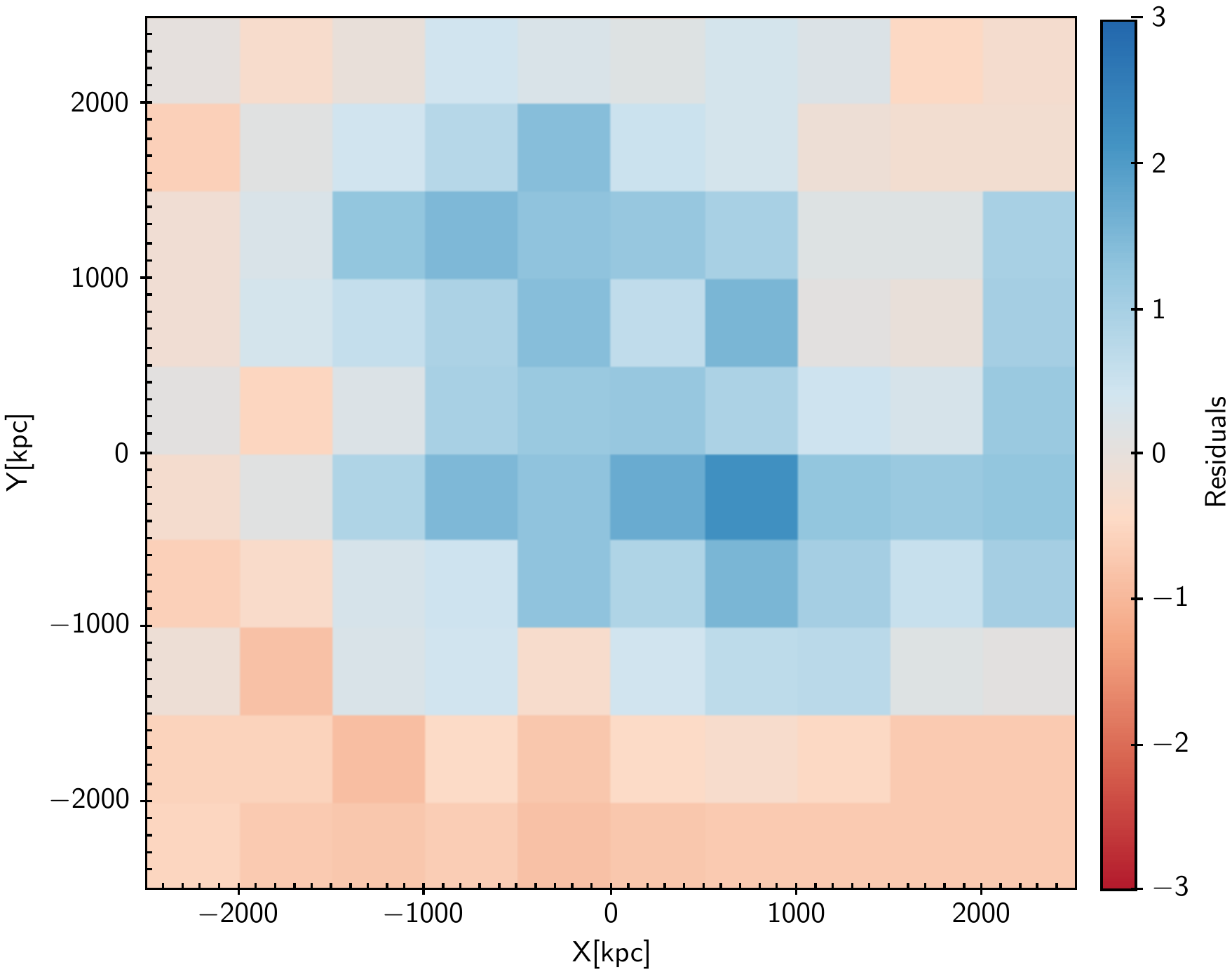}
    \caption{Residuals of least-squares fitting of density averaged for all bins of $Z$ that we use. Left: Residuals of the boxy bulge. Right: Residuals of the X-shaped bulge.}\label{res}
\end{figure*}

\section{Analysis}
We followed the analysis of \cite{Lop17}, who carried out an analysis to determine the distances of Mira variables with $\log P$(days)$<2.6$ within 58 deg$^2$ in the off-plane bulge. The distances can be calculated using the period-luminosity relationship \citep{whitelock,catchpole}

\begin{eqnarray}\label{lambda}
M_K&=&(-3.51 \pm 0.20) \lambda - (7.25 \pm 0.07)~, \\
\lambda &=& \mathrm{log_{10}}[P(\mathrm{days})]-2.38 \label{lambda2}
\end{eqnarray}
and the K-band extinction $A_K = 0.11 A_V$. We determined the extinction $A_V$ using the \cite{schlegel} maps.
Then, we calculated the apparent magnitude as

\begin{equation}
m_K=m_I-(I-K)
,\end{equation}
where $m_I$ is the apparent magnitude in the I filter and $(I-K)$ was determined by \cite{Lop17} as $(I-K)=3.96+3.69 \lambda+10.33 \lambda^2+10.98 \lambda^3$ by calibrating Two Micron All Sky Survey and OGLE data. Finally, for the distance we can use the expression

\begin{equation}
r(m_K)=10^{[m_K-M_K+5]/5}~,
\end{equation}
where $m_K$ is the extinction-corrected K-band magnitude.

The average age of these Mira variables is 9 Gyr, approximately the age of the red-clump samples. The error in the absolute magnitude determination is less than 0.1 mag, which can be neglected.
We repeated the analysis for a much wider coverage, the whole bulge off-plane area of $\approx 250$ deg$^2$. As a result we obtained not only the distribution along a few lines of sight, but also a complete 3D distribution of the stars in the whole
off-plane bulge. We assumed a galactocentric distance of 8.2 kpc and a position
of the Sun over the plane of +17 pc \citep{Kar17}.
We selected only the stars within $|X|\le 2750$ pc, $|Y|\le 2750$ pc, and 700 pc$\le |Z|<1500$ pc for a total of 2\,078 stars, which we divided into bins with sizes $\Delta X=500$ pc, $\Delta Y=500$ pc, and $\Delta Z=200$ pc. The density maps are plotted in Fig. \ref{Fig:slices}.

For $\log P$(days)$<2.6$ Mira variables, the absolute magnitude, $\overline{M_I}$ (averaged over the whole period), 
is lower than -2.6, so for a maximum heliocentric distance of 12 kpc the apparent magnitude, $\overline{m_I}$, is lower than $12.8+A_I$, where $A_I$ is the extinction along the line of sight. The maximum amplitude of the variation
within the period for these Mira stars in bulge fields is lower than $\approx 3$ magnitudes \citep[Fig. 8]{Iwa22}. The limiting magnitude is $m_{I,{\rm max}}=19.0$ for $|b|>5^\circ $ \citep{Uda15}, although for very crowded fields at lower latitudes it can be lower (for instance, a value of $m_{I,{\rm max}}=18.5$ at $\ell =1^\circ $, $b=-2^\circ $ was estimated by \citealt{Uda15}). This means that the Mira variables are always below the limiting
magnitude, provided that $A_I<3$ mag. In our range of Galactic latitudes, $b$, this is the case \citep{sumi_ext,Lop17}. We note, however, that
if we wanted to explore lower $b$ regions (lower $|Z|$), we would lose an important fraction of stars, especially at larger distances. 

\subsection*{Fits of the density}
To fit the data, we used the boxy bulge model from \cite{martin05} and the X-shaped bulge model from \cite{wegg} following the parametrization given by \cite{Lop17}:

\begin{equation}
\rho_{Boxy}(x,y,z)=\rho_0~\mathrm{exp}\left(-\frac{\left(x^4+\left(\frac{y}{0.5}\right)^4+\left(\frac{z}{0.4}\right)^4\right)^{1/4}}{740 ~\mathrm{pc}} \right) \hspace*{-5cm}
\end{equation}

\begin{eqnarray}
\rho_{X-shape}(x,y,z)&=&\rho_0 ~\mathrm{exp} \left(-\frac{s_1}{700~\mathrm{pc}}\right)\mathrm{exp} \left(-\frac{\lvert z \rvert}{322~\mathrm{pc}}\right) \nonumber \\
&\times&\left[1+3~ \mathrm{exp}\left(-\left(\frac{s_2}{1000~\mathrm{pc}}\right)^2\right) \right. \nonumber \\
&+& \left. \vphantom 3~\mathrm{exp} \left(-\left(\frac{s_3}{1000~\mathrm{pc}}\right)^2\right)\right]~, \\
s_1&=& \mathrm{Max}\left[2100~\mathrm{pc}, \sqrt{x^2+\left(\frac{y}{0.7}\right)^2} \right]~, \nonumber \\
s_2&=& \sqrt{(x-1.5z)^2+y^2}~, \nonumber \\
s_2&=& \sqrt{(x+1.5z)^2+y^2}~. \nonumber
\end{eqnarray}
In Fig. \ref{dens} we plot densities of both models compared with the data. We fitted the density maps with only one free parameter, the amplitude $\rho_0$, using the usual least-squares fitting:

\begin{equation}\label{chi2}
\chi^2=\sum\limits_{i=1}^{N} \frac{|\rho_{i,\mathrm{model}}-\rho_{i,\mathrm{data}}|^2}{\sigma_i ^2}.
\end{equation}
We did not vary the rest of the free parameters, as this sample is not appropriate for fine-tuning the parameters of the models. Fitting the data with the least-squares method and minimizing reduced $\chi_r^2\equiv \frac{\chi ^2}{N-1}$ leads to a minimal $\chi_r^2=1.04$ for the boxy bulge model and $\chi_r^2=1.22$ for the X-shaped bulge. Since our sample has 967 degrees of freedom, we used the $\chi^2$ to calculate the probability, which in the case of a boxy bulge is $p=0.19$, whereas in the case of an X-shaped bulge we obtain $p=2.85 \cdot 10^{-6}$. Thus, the boxy bulge fits the density distribution significantly better. For illustration, in Fig. \ref{res} we plot residuals of the fits averaged for every $Z$, calculated as

\begin{equation}
\mathrm{Residuals}=f_i=\frac{\rho_{\mathrm{data}}-\rho_{\mathrm{model}}}{\sigma}~, 
\end{equation}
where $\sigma$ stands for the error of data.

We also experimented with changing the fixed parameters of the models to free parameters to explore if this improves the fits. In both cases, the free parameters are almost identical to the fixed values and the $\chi_r^2$ of the fits improved only negligibly. Therefore, we kept the parameters constant and kept only the amplitude as a free parameter.

As the pulsation period is related to the age of the star with the following relationship \citep{Lop17},
        
\begin{equation}
\mathrm{age(Gyr)}\sim 12-44\lambda+43\lambda^2
,\end{equation}
with $\lambda$ defined by Eq. (\ref{lambda2}), we were able to divide the stars into three groups based on age: $\log P < 2.38$, $2.38 < \log P < 2.53$, and $2.53 < \log P < 2.6$, which approximately correspond to ages $\gtrsim 11.6$ Gyr, 6.4-11.6 Gyr, and $\lesssim 6.4$ Gyr. We fitted each of the populations separately, but in each case $\chi_r$ is significantly lower than 1, and therefore we cannot exclude any model. We conclude that we do not have enough stars of each age to be able to be make separate fits.

In order to improve the fits, we also considered a modification of Eq. (\ref{chi2}) for bins with zero density, where we use
\begin{equation}
f_i(\rho _{\rm data}=0)=-0.716+\sqrt{0.716^2+2.2239\,\mu},
\end{equation}
where $\mu$ is the theoretical expected number of stars per bin. This expression is derived by using $\chi^2=\Sigma f_i^2$, where $f_i$ is the number of sigmas of deviation of an observed point with respect to the theoretical prediction, and calibrating the relationship of $f_i$ with a distribution of probabilities, $P_i$, in a Gaussian distribution, which gives $log_{10}P_i=-0.274f_i-0.194f_i^2$.
Hence, for a Poissonian distribution in pixels where zero stars are detected and the probability is $P_i=exp(-\mu)$, we get the above expression.
This modification leads to a small improvement in the fits for the whole dataset, where we obtain minimal $\chi_r^2=0.99$ for the boxy bulge model and $\chi_r^2=1.19$ for the X-shaped bulge. However, for the individual datasets separated by age, we do not see significant improvement: all values of $\chi_r^2$ are well below 1. Therefore, we consider this modification in the calculation to be $\chi_r^2$ negligible.

                
                
                

\section{Conclusion}
We used the recent OGLE data of Mira variables to analyse the shape of the bulge. We derived density maps of the bulge in an area far from the plane (700 pc$\le |Z|<1500$ pc), which we fitted with density models of a boxy bulge and an X-shaped bulge. Based on least-squares fitting, we calculated the probability of a boxy bulge matching the data to be $p=0.19$, whereas the probability of an X-shaped bulge is $p=2.85 \cdot 10^{-6}$ (equivalent to a 4.7$\sigma $ event). 
Therefore, the boxy model is more fitting for the shape of the bulge, although we cannot completely exclude any model based on this result. We also separated the stars based on age and tried to analyse each population separately, but we lack a sufficient number of stars to be able to make separate fits.
Improving this result only with Mira variables is not possible, since the whole bulge was already covered.\ However, complementing this dataset with other stellar populations may constrain the possibilities for the morphology of the bulge even more.

\begin{acknowledgements}
The authors were supported by the grant PGC-2018-102249-B-100 of the Spanish Ministry of Economy and Competitiveness (MINECO).
\end{acknowledgements}

\bibliographystyle{aa} 
\bibliography{bulge_lang_corr}

\end{document}